\documentclass[12pt] {article}
\usepackage{psfrag}
\usepackage{graphicx}
\usepackage{latexsym,amsfonts}
\usepackage{amssymb}
\begin{document}
\setcounter{page}{1}
\def\theequation{\arabic{section}.\arabic{equation}}
\def\theequation{\thesection.\arabic{equation}}
\setcounter{section}{0}

\title{Is the energy density of the ground state of the sine--Gordon
model unbounded from below for $\beta^2 > 8\pi$\,?}

\author{M. Faber\thanks{E--mail: faber@kph.tuwien.ac.at, Tel.:
+43--1--58801--14261, Fax: +43--1--5864203} ~~and~
A. N. Ivanov\thanks{E--mail: ivanov@kph.tuwien.ac.at, Tel.:
+43--1--58801--14261, Fax: +43--1--5864203}~\thanks{Permanent Address:
State Technical University, Department of Nuclear Physics, 195251
St. Petersburg, Russian Federation}}

\date{\today}

\maketitle
\vspace{-0.5in}
\begin{center}
{\it Atominstitut der \"Osterreichischen Universit\"aten,
Arbeitsbereich Kernphysik und Nukleare Astrophysik, Technische
Universit\"at Wien, \\ Wiedner Hauptstr. 8-10, A-1040 Wien,
\"Osterreich }
\end{center}

\begin{center}
\begin{abstract}
We discuss Coleman's theorem concerning the energy density of the
ground state of the sine--Gordon model proved in Phys. Rev. D {\bf
11}, 2088 (1975). According to this theorem the energy density of the
ground state of the sine--Gordon model should be unbounded from below
for coupling constants $\beta^2 > 8\pi$. The consequence of this
theorem would be the non--existence of the quantum ground state of the
sine--Gordon model for $\beta^2 > 8\pi$. We show that the energy
density of the ground state in the sine--Gordon model is bounded from
below even for $\beta^2 > 8\pi$. This result is discussed in relation
to Coleman's theorem (Comm. Math. Phys. {\bf 31}, 259 (1973)),
particle mass spectra and soliton--soliton scattering in the
sine--Gordon model.
\end{abstract}
\end{center}

\newpage

\section{Introduction}
\setcounter{equation}{0}

\hspace{0.2in} As has been shown in Refs.\cite{FI1,FI3} the massless
Thirring model is unstable under spontaneous breaking of chiral
$U(1)\times U(1)$ symmetry. The non--perturbative phase of
spontaneously broken chiral symmetry is described by a ground state
wave function of BCS--type \cite{FI1}.

The Lagrangian of the massless Thirring model is given by
\cite{FI1}--\cite{WT1}
\begin{eqnarray}\label{label1.1}
{\cal L}_{\rm Th}(x) = \bar{\psi}(x)i\gamma^{\mu}\partial_{\mu}\psi(x)
- \frac{1}{2}\,g\,\bar{\psi}(x)\gamma^{\mu}\psi(x) \bar{\psi}(x)
\gamma_{\mu}\psi(x) + \sigma(x)\,\bar{\psi}(x)\psi(x),
\end{eqnarray}
where $\sigma(x)$ is an external source of the scalar density
$\bar{\psi}(x)\psi(x)$ of the Thirring fermion fields and $g$ is the
coupling constant, which we treat in the attractive case. For
$\sigma(x) = - m$ \cite{FI3}, where $m$ can be interpreted as a mass
of Thirring fermion fields, the Thirring model (\ref{label1.1})
bosonizes to the sine--Gordon model with the Lagrangian \cite{FI1,FI3}
\begin{eqnarray}\label{label1.2}
{\cal L}_{\rm SG}(x) =
\frac{1}{2}\partial_{\mu}\vartheta(x)\partial^{\mu}\vartheta(x) +
\frac{\alpha_0}{\beta^2}\,(\cos\beta\vartheta(x) - 1),
\end{eqnarray}
where $\alpha_0$ and $\beta$ are positive parameters
\cite{FI1,FI3,SC75}. The parameter $\alpha_0$ has the meaning of a
squared mass of the quantum of the sine--Gordon field
\begin{eqnarray}\label{label1.3}
{\cal L}_{\rm SG}(x) &=&
\frac{1}{2}\partial_{\mu}\vartheta(x)\partial^{\mu}\vartheta(x) +
\frac{\alpha_0}{\beta^2}\,(\cos\beta\vartheta(x) - 1) =\nonumber\\ &=&
\frac{1}{2}\partial_{\mu}\vartheta(x)\partial^{\mu}\vartheta(x) -
\frac{1}{2}\,\alpha_0\,\vartheta^2(x) +
\frac{1}{4!}\,\alpha_0\beta^2\,\vartheta^4(x) + \ldots
\end{eqnarray}
and $\beta$ is a coupling constant. For the Thirring fermion fields
quantized in the chirally broken phase the coupling constants $g$ and
$\beta$ are related by \cite{FI1}
\begin{eqnarray}\label{label1.4}
\frac{8\pi}{\beta^2} = 1 - e^{\textstyle - 2\pi/g}.
\end{eqnarray}
The direct consequence of this relation is that $\beta^2 > 8\pi$. As
has been discussed in \cite{FI1}, the relation $\beta^2 > 8\pi$ leads
to a 1+1--dimensional world populated mainly by soliton and
antisoliton states \cite{FI1}, which are classical solutions of the
equations of motion of the sine--Gordon model (\ref{label1.2})
\begin{eqnarray}\label{label1.5}
\Box\vartheta(x)+ \frac{\alpha_0}{\beta}\,\sin\beta\vartheta(x) = 0
\end{eqnarray}
regardless of the value of the coupling constant $\beta$. It is
well--known that there exists an infinite set of dynamical
many--soliton solutions of (\ref{label1.5}) which are collective
excitations of the sine--Gordon field \cite{CR84}.

As an example, the one--soliton and one--antisoliton solutions
$\vartheta_{\rm s}(x^0,x^1)$ and $\vartheta_{\bar{{\rm s}}}(x^0,x^1)$
\begin{eqnarray}\label{label1.6}
\vartheta_{\rm s}(x^0,x^1) &=& \frac{4}{\beta}\,\arctan(\exp\,
(+\sqrt{\alpha_0}\,\gamma(x^1 - ux^0)),\nonumber\\
\vartheta_{\bar{{\rm s}}}(x^0,x^1) &=& \frac{4}{\beta}\,\arctan(\exp\,
(-\sqrt{\alpha_0}\,\gamma(x^1 - u x^0)),
\end{eqnarray}
where $u$ is their velocity and $\gamma = 1/\sqrt{1 - u^2}$ is the
Lorentz factor, have a finite classical mass, $M_{\rm s} = M_{\bar{\rm
s}} = 8\sqrt{\alpha_0}/\beta^2$, and are not related to the quantum
ground state of the sine--Gordon model.

In his pioneering paper \cite{SC75} Coleman has proved the equivalence
between the massive Thirring model and the sine--Gordon model. A
lateral result of Coleman's paper \cite{SC75} was the proof of the
theorem asserting that for $\beta^2 > 8\pi$ the energy density of the
sine--Gordon model is unbounded from below. Due to this Coleman
argued: ``{\it The theory has no ground state, and is physically
nonsensical.}'' \cite{SC75}.  In this paper we discuss critically this
theorem of Coleman and show that the energy of the ground state of the
sine--Gordon model is bounded even for $\beta^2 > 8\pi$.

The paper is organized as follows. In Section 2 we repeat Coleman's
derivation of the theorem asserting the non--existence of the ground
state in the sine--Gordon model for $\beta^2 > 8\pi$ and accentuate
those places where we do not agree with Coleman. We modify Coleman's
derivation and get a bounded energy density for the ground state of
the sine--Gordon model for $\beta^2 > 8\pi$. In Section 3 we adduce
the explicit calculation of the energy density for the ground state of
the sine--Gordon model using the path--integral approach. In Sections
4, 5 and 6 we discuss the relation of the constraint on the coupling
constants $\beta^2 > 8\pi$ to (i) Coleman's theorem, asserting the
non--existence Goldstone bosons in 1+1--dimensional quantum field
theories, to (ii) particle mass spectra of the sine--Gordon model and
to (iii) soliton--soliton scattering.  In the Conclusion we discuss
the obtained results. In the Appendix we follow \cite{FI3} and
evaluate the generating functional of Green functions in the
sine--Gordon model and demonstrate the infrared stability and
non--perturbative renormalizability of this model.

\section{Coleman's proof of the theorem on the unbounded vacuum energy 
density for $\beta^2 > 8\pi$}
\setcounter{equation}{0}

\hspace{0.2in} According to the Lagrangian (\ref{label1.2}) the
Hamiltonian of the sine--Gordon model should be equal to
\begin{eqnarray}\label{label2.1}
{\cal H}_{\rm SG}(x) = \frac{1}{2}\,\Pi^2(x)
+\frac{1}{2}\Bigg(\frac{\partial \vartheta(x)}{\partial
x^1}\Bigg)^{\!2}\! - \frac{\alpha_0}{\beta^2}\,(\cos\beta\vartheta(x)
- 1),
\end{eqnarray}
where $\Pi(x) =\dot{\vartheta}(x)$ is the conjugate momentum of the
$\vartheta$--field.  Following Coleman \cite{SC75} we transcribe the
Hamiltonian (\ref{label2.1}) into the form
\begin{eqnarray}\label{label2.2}
{\cal H}_{\rm SG}(x) = \frac{1}{2}\,\Pi^2(x)
+\frac{1}{2}\,\Bigg(\frac{\partial \vartheta(x)}{\partial
x^1}\Bigg)^{\!2} - \frac{\alpha_0}{\beta^2}\,\cos\beta\vartheta(x) -
\gamma_0,
\end{eqnarray}
where $\gamma_0$ is an arbitrary constant, which is equal to
\begin{eqnarray}\label{label2.3} 
\gamma_0 = - \frac{\alpha_0}{\beta^2},
\end{eqnarray}
if the minimum of the classical potential energy is normalized to zero
\cite{SC75}.

The aim of this section is two--fold corresponding to two scenarios of
the evolution of the sine--Gordon field. In the first scenario the
parameter $\gamma_0$ is arbitrary and additively renormalizable, as
has been assumed by Coleman. We show that in this scenario the ground
state of the sine--Gordon model suffers from an {\it infrared
disaster}. In the infrared limit the renormalized energy density of
the ground state of the sine--Gordon model is equal to negative
infinity at any coupling constant $\beta$. This corresponds to the
non--existence of the sine--Gordon model. Hence, Coleman's proof, when
analysed with respect to the infrared stability of the sine--Gordon
model, leads to the suppression of the sine--Gordon model as quantum
field theory.  In the second scenario the parameter $\gamma_0$ is
fixed to the value (\ref{label2.3}) that normalizes the potential
energy to zero. In this case the parameter $\gamma_0$ is not
additively renormalizable. This results in the energy density of the
ground state of the sine--Gordon model to be (i) positive--definite in
the infrared limit and (ii) stable for any coupling constant $\beta$
even if $\beta^2 > 8\pi$.

First we analyse the stability of the sine--Gordon model following the
scenario when $\gamma_0$ is an additively renormalizable
parameter. Introducing the infrared scale $\mu$, which should be
finally taken in the limit $\mu \to 0$, we can redefine the
interaction term in the Hamiltonian (\ref{label2.2}) as follows
\cite{SC75}
\begin{eqnarray}\label{label2.4}
\cos\beta\vartheta(x) = \Bigg(\frac{\mu^2}{\Lambda^2}\Bigg)^{\textstyle
\beta^2/8\pi}:\!\cos\beta\vartheta(x)\!:_{\mu},
\end{eqnarray}
where the symbol $:\ldots:_{\mu}$ means normal ordering at the scale
$\mu$ and $\Lambda$ is the ultra--violet cut--off. The expression
(\ref{label2.4}) is a trivial consequence of the perturbative
derivation of the vacuum expectation value of the operator
$\cos\beta\vartheta(x)$ 
\begin{eqnarray}\label{label2.5}
\langle 0|\cos\beta\vartheta(x)|0\rangle =e^{\textstyle -
\frac{1}{2}\,\beta^2\,D^{(+)}(0;\mu)} =
\Bigg(\frac{\mu^2}{\Lambda^2}\Bigg)^{\textstyle \beta^2/8\pi},
\end{eqnarray}
where $D^{(+)}(x;\mu)$ is the two--point Wightman function defined by
\begin{eqnarray}\label{label2.6}
D^{(+)}(x; \mu) = \langle
0|\vartheta(x)\vartheta(0)|0\rangle =
\frac{1}{2\pi}\int^{\infty}_{-\infty}\frac{dk^1}{2k^0}\,e^{\textstyle
-\,i\,k\cdot x} = - \frac{1}{4\pi}\,{\ell n}[-\mu^2x^2 +
i\,0\cdot\varepsilon(x^0)].
\end{eqnarray}
For $x = (x^0,x^1) = 0$ the two--point Wightman function is
regularized by the ultra--violet cut--off $\Lambda$, $|k^1| \le
\Lambda$, and reads
\begin{eqnarray}\label{label2.7}
D^{(+)}(0; \mu) = \frac{1}{4\pi}\,{\ell
n}\Bigg(\frac{\Lambda^2}{\mu^2}\Bigg).
\end{eqnarray}
Since the vacuum expectation value of the normal--ordered operator
$:\!\cos\beta\vartheta(x)\!:$ is unity, $\langle
0|:\!\cos\beta\vartheta(x)\!:|0\rangle = 1$, relation (\ref{label2.5})
can be represented in the operator form (\ref{label2.4}). Of course,
the same result can be obtained by considering the $\vartheta$--field
as a free field and applying Wick's theorem \cite{SC75,GW50}.

Assuming multiplicative renormalizability of the sine--Gordon model
Coleman (i) introduces the renormalized constant $\alpha$ determined
by
\begin{eqnarray}\label{label2.8}
\alpha = \alpha_0\, \Bigg(\frac{\mu^2}{\Lambda^2}\Bigg)^{\textstyle
\beta^2/8\pi}
\end{eqnarray}
and (ii) changes the scale of the normal ordering $\mu \to M$
according to the recipe \cite{SC75}
\begin{eqnarray}\label{label2.9}
:\!\cos\beta\vartheta(x)\!:_{\mu}\to
\Bigg(\frac{M^2}{\mu^2}\Bigg)^{\textstyle
\beta^2/8\pi}:\!\cos\beta\vartheta(x)\!:_M.
\end{eqnarray}
As a result the interaction term of the Hamiltonian of the
sine--Gordon model acquires the form
\begin{eqnarray}\label{label2.10}
{\cal H}^{int}_{\rm SG}(x) = -
\frac{\alpha}{\beta^2}\,\Bigg(\frac{M^2}{\mu^2}\Bigg)^{\textstyle
\beta^2/8\pi}:\!\cos\beta\vartheta(x)\!:_M,
\end{eqnarray}
where the parameter $\alpha$ is related to the {\it bare} parameter
$\alpha_0$ by equation (\ref{label2.8}). This completes the
redefinition of the interaction part of the Hamiltonian
(\ref{label2.2}).

Now according to Coleman we rewrite the free part of the Hamiltonian
(\ref{label2.2}) as follows
\begin{eqnarray}\label{label2.11}
{\cal H}^{(0)}_{\rm SG}(x) = \frac{1}{2}\,\Pi^2(x)
+\frac{1}{2}\,\Bigg(\frac{\partial \vartheta(x)}{\partial
x^1}\Bigg)^{\!2} = :\!\frac{1}{2}\,\Pi^2(x)
+\frac{1}{2}\,\Bigg(\frac{\partial \vartheta(x)}{\partial
x^1}\Bigg)^{\!2}\!:_{\mu} + {\cal E}_0(\mu),
\end{eqnarray}
where ${\cal E}_0(\mu)$ is equal to \cite{SC75}
\begin{eqnarray}\label{label2.12}
{\cal E}_0(\mu) =
\int^{\infty}_{-\infty}\frac{dk^1}{8\pi}\,\frac{2(k^1)^2 +
\mu^2}{\sqrt{(k^1)^2 + \mu^2}}.
\end{eqnarray}
The regularized version of ${\cal E}_0(\mu)$ reads
\begin{eqnarray}\label{label2.13}
{\cal E}_0(\Lambda, \mu) =
\int^{\Lambda}_{-\Lambda}\frac{dk^1}{8\pi}\,\frac{2(k^1)^2 +
\mu^2}{\sqrt{(k^1)^2 + \mu^2}} = \frac{\Lambda^2}{4\pi}\,\sqrt{1 +
\frac{\mu^2}{\Lambda^2}} = \frac{\Lambda^2}{4\pi} + \frac{\mu^2}{8\pi}
+ O\Big(\frac{\mu^4}{\Lambda^2}\Big).
\end{eqnarray}
The appearance of ${\cal E}_0(\Lambda, \mu)$ can be easily justified
using the expansions of the field $\vartheta(x)$ and the conjugate
momentum $\Pi(x)$ into plane waves \cite{SC75,FI2} 
\begin{eqnarray}\label{label2.14}
\vartheta(x) &=&\frac{1}{2\pi}
\int^{\infty}_{-\infty}\frac{dk^1}{2k^0}\, \Big(a(k^1)\,e^{\textstyle
-i\,k\cdot x} + a^{\dagger}(k^1)\,e^{\textstyle i\,k\cdot
x}\Big),\nonumber\\ \Pi(x) &=&\frac{1}{2\pi}
\int^{\infty}_{-\infty}\frac{dk^1}{2i}\, \Big(a(k^1)\,e^{\textstyle
-i\,k\cdot x} - a^{\dagger}(k^1)\,e^{\textstyle i\,k\cdot x}\Big),
\end{eqnarray}
where $k^0 = \sqrt{(k^1) + \mu^2}$, $a(k^1)$ and $a^{\dagger}(k^1)$
are annihilation and creation operators obeying the standard
commutation relation
\begin{eqnarray}\label{label2.15}
[a(k^1), a^{\dagger}(q^1)] = (2\pi)\,2k^0\,\delta(k^1 - q^1).
\end{eqnarray}
Assuming additive renormalizability of the parameter $\gamma_0$
Coleman defines the renormalized parameter $\gamma$
\begin{eqnarray}\label{label2.16}
\gamma = \gamma_0 + {\cal E}_0(\mu).
\end{eqnarray}
and redefines the free part of the Hamiltonian (\ref{label2.2}) as
follows
\begin{eqnarray}\label{label2.17}
{\cal H}^{(0)}_{\rm SG}(x) - \gamma_0 = :\!{\cal H}^{(0)}_{\rm
SG}(x)\!:_{\mu} - \gamma.
\end{eqnarray}
After this set of transformations Coleman asserts that: ``{\it
Assembling all this, we find the cut--off independent form of the
Hamiltonian density}'':
\begin{eqnarray}\label{label2.18}
{\cal H}_{\rm SG}(x) = :\!\frac{1}{2}\,\Pi^2(x)
+\frac{1}{2}\,\Bigg(\frac{\partial \vartheta(x)}{\partial
x^1}\Bigg)^{\!2} -
\frac{\alpha}{\beta^2}\,\cos\beta\vartheta(x)\!:_{\mu} - \gamma,
\end{eqnarray}
where all parameters $\alpha$, $\beta$ and $\gamma$ are finite.

Changing then the scale of the normal--ordering $\mu \to M$
\cite{SC75} 
\begin{eqnarray}\label{label2.19}
:\!{\cal H}^{(0)}_{\rm SG}(x)\!:_{\mu} \to :\!{\cal
H}^{(0)}_{\rm SG}(x)\!:_M + {\cal E}_0(M) - {\cal
E}_0(\mu) = :\!{\cal H}^{(0)}_{\rm SG}(x)\!:_M +
\frac{1}{8\pi}\,(M^2 - \mu^2)
\end{eqnarray}
Coleman arrives at the Hamiltonian
\begin{eqnarray}\label{label2.20}
{\cal H}_{\rm SG}(x) &=& :\frac{1}{2}\,\Pi^2(x) +
\frac{1}{2}\,\Bigg(\frac{\partial \vartheta(x)}{\partial
x^1}\Bigg)^{\!2}:_M - \frac{\alpha}{\beta^2}
\Big(\frac{M^2}{\mu^2}\Big)^{\textstyle\beta^2/8\pi}:\cos\beta\vartheta(x)\!:_M\nonumber\\
&&+ \frac{1}{8\pi}\,(M^2 - \mu^2) - \gamma.
\end{eqnarray}
The renormalized energy density of the ground state is equal to
\begin{eqnarray}\label{label2.21}
{\cal E}_{\rm vac}(M) = - \frac{\alpha}{\beta^2}
\Big(\frac{M^2}{\mu^2}\Big)^{\textstyle\beta^2/8\pi} +
\frac{1}{8\pi}\,(M^2 - \mu^2) - \gamma.
\end{eqnarray}
This is Eq.(3.7) of Ref.\cite{SC75}. The renormalized energy density
${\cal E}_{\rm vac}(M)$ (\ref{label2.21}) depends explicitly on the
infrared cut--off $\mu$, which should be taken in the limit $\mu \to
0$, whereas all parameters $\alpha$, $\beta$ and $\gamma$ are kept
finite. Before one analyses the behaviour of the renormalized energy
density in the limit $M \to \infty$, one has to take the limit $\mu
\to 0$. Taking the limit $\mu \to 0$ one gets the renormalized energy
density (\ref{label2.21}) equal to negative infinity for any finite
scale $M$ and any coupling constant $\beta \neq 0$. This makes the
sine--Gordon model to be an extremely ill--defined quantum field
theory and {\it nonsensical} for any coupling constant $\beta \neq
0$. This contradicts the infrared stability of the sine--Gordon model
(see, for example, Appendix to this paper) and gives no constraints on
the value of the coupling constant like $\beta^2 < 8\pi$.

Such an {\it infrared disaster} is a consequence of two of Coleman's
assumptions, the finiteness of the parameter $\alpha$ and the additive
renormalizability of the parameter $\gamma$.  The finiteness of the
parameter $\alpha$ in (\ref{label2.8}) is questionable, since this
entails the infinity of the parameter $\alpha_0$ in the infrared limit
$\mu \to 0$. The former is not really true. Indeed, as we have shown
(see the Appendix to this paper) the sine--Gordon model is
non--singular in the infrared limit ({\rm A}.12) and the correlation
functions of the sine--Gordon model are finite in this
limit. Moreover, if $\alpha_0$ would be infinite in the infrared limit
the soliton solutions of the sine--Gordon model like (\ref{label1.6})
would not exist. Hence, there is no physical reason for the parameter
$\alpha_0$ to be infinite at $\mu \to 0$.

Thus, Coleman's scenario of the evolution of the sine--Gordon field
with an arbitrary and additively renormalizable parameter $\gamma_0$
leads to the {\it infrared disaster} of the sine--Gordon model and
makes no constraints on the value of the coupling constant $\beta$.

Now let us analyse another scenario of the evolution of the
sine--Gordon field with a potential energy normalized to zero. In this
case the parameter $\gamma_0$ is fixed to the value (\ref{label2.3})
and after the renormalization of the parameter $\alpha_0$
(\ref{label2.8}) one should get the renormalized $\gamma$, i.e.
\begin{eqnarray}\label{label2.22} 
\gamma = -
\frac{\alpha}{\beta^2}\Bigg(\frac{\Lambda^2}{\mu^2}\Bigg)^{\textstyle
\beta^2/8\pi}.
\end{eqnarray}
This yields the Hamiltonian (\ref{label2.18}) depending
explicitly on the ultra--violet cut--off $\Lambda$
\begin{eqnarray}\label{label2.23}
\hspace{-0.3in}{\cal H}_{\rm SG}(x) &=& :\!\frac{1}{2}\,\Pi^2(x)
+\frac{1}{2}\,\Bigg(\frac{\partial \vartheta(x)}{\partial
x^1}\Bigg)^{\!2}:_{\mu} -
\frac{\alpha}{\beta^2}:\cos\beta\vartheta(x)\!:_{\mu}\nonumber\\ && +
\frac{\alpha}{\beta^2}\Bigg(\frac{\Lambda^2}{\mu^2}\Bigg)^{\textstyle
\beta^2/8\pi} + {\cal E}_0(\Lambda, \mu).
\end{eqnarray}
Since unlike Coleman through the parameter $\gamma_0$ the Hamiltonian
depends of the ultra--violet cut--off $\Lambda$, one does not need to
remove the ultra--violet divergence of ${\cal E}_0(\Lambda, \mu)$
appearing due to the normal ordering of the kinetic term.

Following then Coleman and changing the scale of the normal ordering
$\mu \to M$ we arrive at the Hamiltonian
\begin{eqnarray}\label{label2.24}
{\cal H}_{\rm SG}(x) &=& :\!\frac{1}{2}\,\Pi^2(x)
+\frac{1}{2}\,\Bigg(\frac{\partial \vartheta(x)}{\partial
x^1}\Bigg)^{\!2}:_M -
\frac{\alpha}{\beta^2}\,\Bigg(\frac{M^2}{\mu^2}\Bigg)^{\textstyle
\beta^2/8\pi}\!:\cos\beta\vartheta(x):_M\nonumber\\ &&
+\frac{\alpha}{\beta^2}\Bigg(\frac{\Lambda^2}{\mu^2}\Bigg)^{\textstyle
\beta^2/8\pi}\! + {\cal E}_0(\Lambda, M).
\end{eqnarray}
The vacuum energy density defined by the Hamiltonian (\ref{label2.24})
is equal to
\begin{eqnarray}\label{label2.25}
{\cal E}_{\rm vac}(M) =
\frac{\alpha}{\beta^2}\Bigg(\frac{\Lambda^2}{\mu^2}\Bigg)^{\textstyle
\beta^2/8\pi}\! -
\frac{\alpha}{\beta^2}\,\Bigg(\frac{M^2}{\mu^2}\Bigg)^{\textstyle
\beta^2/8\pi}\! + \frac{\Lambda^2}{4\pi}\,\sqrt{1 +
\frac{M^2}{\Lambda^2}}.
\end{eqnarray}
Due to the first term in the r.h.s. of (\ref{label2.25}), which is
absent in Coleman's expression given by Eq.(3.7) of Ref.\cite{SC75},
the energy density of the ground state of the sine--Gordon model is
positive--definite in the infrared limit. At $M = \Lambda$ the energy
density does not depend on the infrared cut--off and is proportional
to ${\cal E}_{\rm vac}(\Lambda) \sim \Lambda^2$. The problem of this
quadratic ultra--violet divergence can be easily solved taking the
full Hamiltonian (\ref{label2.1}) in the normal--ordered form.

Thus, we argue that the sine--Gordon model with the potential energy
normalized to zero is stable in the infrared limit and the energy of
the ground state can never be negative for arbitrary values of the
coupling constant $\beta$ even for $\beta^2 > 8\pi$.

\section{Vacuum energy density in the sine--Gordon model. 
Non--perturbative calculation} 
\setcounter{equation}{0}

\hspace{0.2in} Using the Lagrangian ${\cal L}_{\rm SG}(x)$ given by
(\ref{label1.2}) one can obtain the Hamilton functional $H(x^0)$ of
the sine--Gordon model
\begin{eqnarray}\label{label3.1}
H(x^0) = \int^{\infty}_{-\infty} dx^1\Bigg\{\frac{1}{2}:\Bigg[\Pi^2(x)
+ \Bigg(\frac{\partial \vartheta(x)}{\partial x^1}\Bigg)^{\!2}\Bigg]:
-\, \frac{\alpha_0}{\beta^2}\,:[\cos\beta\vartheta(x) - 1]:\Bigg\}.
\end{eqnarray}
The first two terms describe the contribution of the kinetic energy
 which should be always taken in the normal--ordered
 form\,\footnote{We carry out the normal ordering at the infrared
 scale $\mu$, which is taken finally in the limit $\mu \to 0$ (see the
 Appendix).}. In quantum field theory the potential energy, given by
 the last two terms in (\ref{label3.1}), should be normal ordered as
 well as the kinetic one. However, below we consider two possibilities
 (i) the potential energy is normal--ordered and (ii) the potential
 energy is not normal--ordered. We will show that in the case of the
 potential energy, taken in the normal--unordered form, the energy of
 the ground state of the sine--Gordon model tends to positive
 infinity.

The energy of the ground state $E_{\rm vac}$ is equal to the vacuum
expectation value of the Hamilton functional $H(x^0)$
\begin{eqnarray}\label{label3.2}
E_{\rm vac} =\langle 0|H(x^0)|0 \rangle = -
\frac{\alpha_0}{\beta^2}\int^{\infty}_{-\infty}
dx^1\,[\langle 0|\cos\beta\vartheta(x)|0 \rangle -
1],
\end{eqnarray}
Since the integrand in (\ref{label3.2}) does not depend on $x$,
instead of the energy of the ground state $E_{\rm vac}$ it is
convenient to treat the vacuum energy density ${\cal E}_{\rm vac}$
defined by
\begin{eqnarray}\label{label3.3}
{\cal E}_{\rm vac} = \lim_{L \to \infty}\frac{E_{\rm
vac}}{L} = - \frac{\alpha_0}{\beta^2}\,[\langle
0|\cos\beta\vartheta(0)|0 \rangle - 1],
\end{eqnarray}
where $L$ is the spatial volume.

(i) If the potential energy is taken in the normal--ordered form, the
energy density ${\cal E}_{\rm vac}$ is equal to zero, $ {\cal E}_{\rm
vac} = 0$, due to $\langle 0|:\cos\beta\vartheta(0):|0 \rangle = 1$ by
definition of the normal ordering.

(ii) In the case of the normal--unordered form of the potential energy
the vacuum expectation value $\langle 0|[\cos\beta\vartheta(0) - 1]|0
\rangle$ is non--zero and can be calculated explicitly. In terms of
the partition function $Z_{\rm SG}[0]$ defined by ({\rm A}.15) the
vacuum expectation value $\langle 0|[\cos\beta\vartheta(0) - 1]|0
\rangle$ reads
\begin{eqnarray}\label{label3.4}
\frac{\alpha_0}{\beta^2}\,[\langle
0|\cos\beta\vartheta(0)|0 \rangle - 1] =\lim_{T,L\to
\infty}\frac{1}{TL}\frac{\alpha_0}{i}\frac{\partial {\ell
n}Z_{\rm SG}[0]}{\partial\alpha_0},
\end{eqnarray}
where $TL$ defines a 1+1--dimensional volume, $\int d^2x = \int
dx^0dx^1 = T L$, at $T, L \to \infty$.

By the renormalization $\alpha_0 \to Z_1\alpha$, where $Z_1$ is a
renormalization constant ({\rm A}.13), we obtain
\begin{eqnarray}\label{label3.5}
\frac{\alpha_0}{\beta^2}\,[\langle
0|\cos\beta\vartheta(0)|0 \rangle - 1] = \lim_{T,L\to
\infty}\frac{1}{TL}\frac{\alpha}{i}\frac{\partial {\ell n}Z_{\rm
SG}[0]}{\partial\alpha},
\end{eqnarray}
Substituting (\ref{label3.5}) in (\ref{label3.3}) we determine the
vacuum energy density ${\cal E}_{\rm vac}$ in terms of the partition
function $Z_{\rm SG}[0]$ as follows
\begin{eqnarray}\label{label3.6}
{\cal E}_{\rm vac} = - \lim_{T,L\to
\infty}\frac{1}{TL}\frac{\alpha}{i}\frac{\partial {\ell n}Z_{\rm
SG}[0]}{\partial\alpha}.
\end{eqnarray}
Due to ({\rm A}.15) the vacuum energy density ${\cal E}_{\rm vac}$ is
equal to
\begin{eqnarray}\label{label3.7}
&&{\cal E}_{\rm vac} = \frac{\alpha}{\beta^2}\,
\Bigg(\frac{\Lambda^2}{M^2}\Bigg)^{\!\textstyle \beta^2/8\pi} +
\frac{2}{i}\lim_{T,L\to
\infty}\frac{1}{TL}\sum^{\infty}_{p=1}\frac{(-1)^p}{p!(p-1)!}\,
\Bigg(\frac{\alpha}{2\beta^2}\Bigg)^{2p} \prod^{p}_{j=1}\int\!\!\!\int
d^2x_j d^2y_j\,\nonumber\\
&&\times\exp\Big\{\frac{\beta^2}{4\pi}\sum^p_{j < k} \Big({\ell n}[ -
M^2(x_j - x_k)^2 + i0] + {\ell n}[ - M^2(y_j - y_k)^2 +
i0]\Big)\nonumber\\ && -
\frac{\beta^2}{4\pi}\sum^p_{j=1}\sum^p_{k=1}{\ell n}[ - M^2(x_j -
y_k)^2 + i0]\Big\}\nonumber\\
&&\times\Bigg[\sum^{\infty}_{q=0}\frac{(-1)^q}{(q!)^2}\,
\Bigg(\frac{\alpha}{2\beta^2}\Bigg)^{2q} \prod^{q}_{j=1}\int\!\!\!\int
d^2x_j d^2y_j\, \exp\Big\{\frac{\beta^2}{4\pi}\sum^q_{j < k}
\Big({\ell n}[ - M^2(x_j - x_k)^2 + i0]\nonumber\\ &&+ {\ell n}[ -
M^2(y_j - y_k)^2 + i0]\Big) -
\frac{\beta^2}{4\pi}\sum^q_{j=1}\sum^q_{k=1}{\ell n}[ - M^2(x_j -
y_k)^2 + i0]\Big\}\Bigg]^{-1},
\end{eqnarray}
where the second term is a ratio of two infinite series and $M$ is a
finite scale. One can show that this term vanishes in the limit $T,L
\to \infty$. For this aim we rewrite the vacuum energy density
(\ref{label3.7}) in the form
\begin{eqnarray}\label{label3.8}
\hspace{-0.3in}&&{\cal E}_{\rm vac} = \frac{\alpha}{\beta^2}\,
\Bigg(\frac{\Lambda^2}{M^2}\Bigg)^{\!\textstyle \beta^2/8\pi} +
\frac{2}{i}\lim_{T,L\to
\infty}\Bigg(\frac{\alpha}{2\beta^2}\Bigg)^2\Bigg[ - \int
\frac{d^2x}{(-M^2x^2 + i0)^{\textstyle \beta^2/4\pi}} +
\frac{1}{2}\Bigg(\frac{\alpha}{2\beta^2}\Bigg)^2\nonumber\\
\hspace{-0.3in}&&\times \int \frac{\displaystyle
d^2xd^2yd^2z\,[(-M^2(x - y)^2 + i0)(-M^2z^2 + i0)]^{\textstyle
\beta^2/4\pi}}{\displaystyle [(-M^2x^2 + i0)(-M^2y^2 + i0)(-M^2(x -
y)^2 + i0)(-M^2(y - z)^2 + i0)]^{\textstyle \beta^2/4\pi}} +
\ldots\Bigg]\nonumber\\ \hspace{-0.3in}&&:\Bigg[1
-\,TL\,\frac{1}{2}\,\Bigg(\frac{\alpha}{2\beta^2}\Bigg)^2 \int
\frac{d^2x}{(-M^2x^2 + i0)^{\textstyle \beta^2/4\pi}} +
\,TL\,\frac{1}{4}\Bigg(\frac{\alpha}{2\beta^2}\Bigg)^4\nonumber\\
\hspace{-0.3in}&&\times\int \frac{\displaystyle d^2xd^2yd^2z\,
[(-M^2(x - y)^2 + i0)(-M^2z^2 + i0)]^{\textstyle
\beta^2/4\pi}}{\displaystyle [(-M^2x^2 + i0)(-M^2y^2 + i0)(-M^2(x -
y)^2 + i0)(-M^2(y - z)^2 + i0)]^{\textstyle \beta^2/4\pi}} +
\ldots\Bigg].\nonumber\\
\hspace{-0.3in}&&
\end{eqnarray}
It is seen that in the limit $T, L \to \infty$ the ratio of the series
is of order $O(1/TL)$. This allows to rewrite (\ref{label3.8}) as
follows
\begin{eqnarray}\label{label3.9}
{\cal E}_{\rm vac} = \frac{\alpha}{\beta^2}\,
\Bigg(\frac{\Lambda^2}{M^2}\Bigg)^{\!\textstyle \beta^2/8\pi} +
\frac{2}{i}\lim_{T,L\to \infty}O\Big(\frac{1}{TL}\Big).
\end{eqnarray}
Hence, in the limit $T, L \to \infty$ the vacuum energy density ${\cal
E}_{\rm vac}$ is defined only by the first term in
(\ref{label3.8}). This gives
\begin{eqnarray}\label{label3.10}
{\cal E}_{\rm vac} = \frac{\alpha}{\beta^2}\,
\Bigg(\frac{\Lambda^2}{M^2}\Bigg)^{\!\textstyle
\beta^2/8\pi}.
\end{eqnarray}
Since the renormalized coupling constant $\alpha$ is finite as well as
the coupling constant $\beta$, in the limit $\Lambda \to \infty$ the
vacuum energy density ${\cal E}_{\rm vac}$ tends to positive infinity
as it is usual for renormalizable quantum field theories with Hamilton
functionals taken in the normal--unordered form.

We would like to remind that Coleman's expression for the energy
density of the ground state of the sine--Gordon model is linear in the
coupling constant $\alpha_0$. Therefore, formally, for the
verification of Coleman's result we can consider only the lowest order
in perturbation theory with respect to the coupling constant
$\alpha_0$. Taking the potential energy in the normal--unordered form
and keeping only the lowest order in the $\alpha_0$ expansion the
vacuum expectation value $\langle 0|\cos\beta\vartheta(0)|0\rangle$
amounts to
\begin{eqnarray}\label{label3.11}
\langle 0|\cos\beta\vartheta(0)|0\rangle = \lim_{\textstyle \mu
\to 0}\Bigg(\frac{\mu^2}{\Lambda^2}\Bigg)^{\textstyle
\beta^2/8\pi} = 0.
\end{eqnarray}
This gives the vacuum energy density (\ref{label3.3}) equal to ${\cal
E}_{\rm vac} = \alpha_0/\beta^2$, which reduces to (\ref{label3.10})
after renormalization $\alpha_0 = \alpha\,Z_1
=\alpha\,(\Lambda^2/M^2)^{\textstyle \beta^2/8\pi}$ with the
renormalization constant $Z_1 = (\Lambda^2/M^2)^{\textstyle
\beta^2/8\pi}$ defined by ({\rm A}.13).

The vacuum energy density (\ref{label3.10}) tends to infinity at
$\Lambda \to \infty$.  Such an infinity can be removed by
normal--ordering. Hence, according to standard conclusions of quantum
field theory the energy of the ground state of the sine--Gordon model
is equal to zero, if the Hamilton functional is taken in the
normal--ordered form.

Within the path--integral approach, where the vacuum energy density of
the sine--Gordon model is defined by the generating functional of
Green functions $Z_{\rm SG}[J]$ for the external source zero, $J =
0$. The energy density of the ground state of the sine--Gordon model
can be set zero normalizing $Z_{\rm SG}[J]$ to unity at $J = 0$,
i.e. $Z_{\rm SG}[0] = 1$.

In the following sections we discuss our result for the ground
state of the sine--Gordon model to be bounded from below for $\beta^2
> 8\pi$ in relation to (i) Coleman's theorem \cite{SC73},
asserting the absence of Goldstone bosons and spontaneously broken
continuous symmetry in quantum field theories in 1+1--dimensional
space--time with Wightman's observables defined on the test functions
from the Schwartz class ${\cal S}(\mathbb{R}^2)$ \cite{AW64}, (ii)
particle mass spectra and (iii) soliton--soliton scattering in the
sine--Gordon model.

\section{Relation to Coleman's theorem:''{\it There are no Goldstone 
Bosons in Two Dimensions}''}
\setcounter{equation}{0}

\hspace{0.2in} The constraint $\beta^2 > 8\pi$ on the coupling
constants $\beta$ appears as a result of the bosonization of the
massless Thirring model with fermion fields quantized in the chirally
broken phase \cite{FI1} and the normalization of the Lagrangian of the
free massless (pseudo)scalar field $\vartheta(x)$ to the standard form
${\cal L}(x) = \frac{1}{2} \partial_{\mu}\vartheta(x)
\partial^{\mu}\vartheta(x)$. Coupling constants $\beta^2 > 8\pi$
define the non--linear response of the free massless (pseudo)scalar
field $\vartheta(x)$ on external sources of Thirring fermion
fields. The wave function of the ground state of the free massless
(pseudo)scalar field has been obtained through the bosonization of the
BCS--type wave function of the ground state of the massless Thirring
model in the chirally broken phase \cite{FI8}. This wave function is
not invariant under chiral transformations, related to the constant
shifts of the free massless (pseudo)scalar field $\vartheta(x) \to
\vartheta(x) + \alpha$, and caused fully by the collective zero--mode
of this field \cite{FI2,FI8,FI9}. The collective zero--mode of the
free massless (pseudo)scalar field $\vartheta(x)$, describing the
motion of the ``center of mass'' of the system, is responsible for the
infrared divergences of the two--point Wightman functions \cite{FI2},
which lead to the vanishing of the generating functional of Green
functions $Z[J]$
\[Z[J] = \int {\cal D}\vartheta\,\exp\Big\{i\int d^2x\,
\Big[\frac{1}{2}\,\partial_{\mu}\vartheta(x)\partial^{\mu}\vartheta(x)
+ \vartheta(x)J(x)\Big]\Big\}\] of the field $\vartheta(x)$, where
$J(x)$ is the external source of this field.

The non--vanishing value of $Z[J]$ can be obtained by the removal of
the collective zero--mode from the spectrum of observable modes. This
can be carried out by the constraint on the external source $\int
d^2x\,J(x) = \tilde{J}(0) = 0$ \cite{FI3,FI2,FI8} (see also ({\rm
A}.3) of the Appendix)\,\footnote{Recall, that the removal of the
collective zero--mode from the spectrum of observable modes has been
discussed by Hasenfratz \cite{PH84} in connection with a correct
formulation of Feynman rules in one and two--dimensional non--linear
$\sigma$--models with $O(N)$ symmetry.}.

As has been pointed out by Wightman \cite{AW64} the quantum field
theory of a free massless (pseudo)scalar field in 1+1--dimensional
space--time does not exist from a mathematical point of view, if
Wightman's observables are defined on the test functions $h(x)$ from
the Schwartz class ${\cal S}(\mathbb{R}^2)$. In this case Wightman's
positive definiteness condition is violated due to infrared
divergences of the two--point Wightman functions
\cite{AW64}. Nevertheless, Wightman has argued that the problem of the
violation of Wightman's positive definiteness condition can be avoided
defining Wightman's observables on the test functions from the Schwartz
class ${\cal S}_0(\mathbb{R}^2) = \{h(x) \in {\cal S}(\mathbb{R}^2);
\tilde{h}(0) = 0\}$, where $\tilde{h}(k)$ is the Fourier transform of
the test function $h(x)$. As has been shown in \cite{FI7} the quantum
field theory of the free massless (pseudo)scalar field with Wightman's
observables defined on the test functions from ${\cal S}_0(\mathbb{R}^2)$
is equivalent to the quantum field theory determined by the generating
functional of Green functions $Z[J]$ with external sources obeying the
constraint $\int d^2x\,J(x) = \tilde{J}(0) = 0$. Since the collective
zero--mode is not induced, such a quantum field theory does not suffer
from infrared divergences of the two--point Wightman functions
\cite{FI2}.

In \cite{SC73} Coleman has reformulated Wightman's ban on the
construction of the quantum field theory of a free massless
(pseudo)scalar field in 1+1--dimensional space--time with Wightman's
observables defined on the test functions $h(x) \in {\cal
S}(\mathbb{R}^2)$ as non--existence of Goldstone bosons, massless
(pseudo)scalar fields, and spontaneously broken continuous symmetry in
1+1--dimensional quantum field theories. The removal of the collective
zero--mode from the system allows to formulate in 1+1--dimensional
space--time a consistent quantum field theory of a free massless
(pseudo)scalar field without infrared divergences. This quantum field
theory is equivalent to Wightman's version of the quantum field theory
of a free massless (pseudo)scalar field with Wightman's observables
defined on the test functions from ${\cal S}_0(\mathbb{R}^2)$. Since
Coleman's theorem concerns only 1+1--dimensional quantum field
theories with Wightman's observables defined on the test functions
from ${\cal S}(\mathbb{R}^2)$ and tells nothing about the absence of
Goldstone bosons and  spontaneous breaking of continuous symmetry in
quantum field theories with Wightman's observables defined on the test
functions from ${\cal S}_0(\mathbb{R}^2)$ \cite{FI8,FI7}, the coupling
constants, obeying the constraint $\beta^2 > 8\pi$, do not contradict
Coleman's theorem \cite{SC73}. This is because such coupling constants
are related to the quantum field theory with Wightman's observables
defined on the test functions from ${\cal S}_0(\mathbb{R}^2)$
\cite{FI8,FI7}.

The sine--Gordon model has been obtained through the bosonization of
the massive Thirring model with fermion fields quantized relative to
the non--perturbative BCS--type superconducting vacuum \cite{FI1}. The
constraint $\beta^2 > 8\pi$ on the coupling constant $\beta$ has
appeared naturally due to the normalization of the kinetic term of the
Lagrangian of the sine--Gordon field to $\frac{1}{2}\,
\partial_{\mu}\vartheta(x) \partial^{\mu}\vartheta(x)$.  Therefore,
the sine--Gordon field $\vartheta(x)$ has inherited all properties of
the free massless (pseudo)scalar field $\vartheta(x)$, bosonizing the
massless Thirring model in the chirally broken phase, which have been
extended by the inclusion of the sine--Gordon interaction. This means
that in our approach the sine--Gordon model is a quantum field theory
of a self--coupled (pseudo)scalar field $\vartheta(x)$ with Wightman's
observables defined on the test functions from the Schwartz class
${\cal S}_0(\mathbb{R}^2)$ (see Appendix).

\section{Particle mass spectra}
\setcounter{equation}{0}

\hspace{0.2in} According to Korepin, Kulish and Faddeev \cite{VK75}
the sine--Gordon model describes three sorts of particle states with
masses: (i) $M_{\rm q} = \sqrt{\alpha_0}$, (ii) $M_{\rm s} =
M_{\bar{\rm s}} = 8\sqrt{\alpha_0}/\beta^2$ and (iii) $M^{(n)}_{\rm
br} = 2M_{\rm s}\sin \nu_n$, where $\nu_n = n\beta^2/16$ with $n =
1,2,\ldots, 8\pi/\beta^2$.

The particles with mass $M_{\rm q} = \sqrt{\alpha_0}$ are the quanta
of the sine--Gordon field in the perturbative regime $\beta^2 \ll
4\pi$, when the potential $V[\vartheta(x)] = (\alpha_0/\beta^2)(1 -
\cos\beta\vartheta(x))$ can be expanded in powers of $\beta^2$,
\begin{eqnarray}\label{label5.1}
V[\vartheta(x)] = 
 \frac{1}{2}\,\alpha_0\,\vartheta^2(x) -
\frac{1}{24}\,\alpha_0\,\beta^2\,\vartheta^4(x) + \ldots\,.
\end{eqnarray}
These quanta are described by the operators of annihilation and
creation in the expansion of the $\vartheta$--field into plane waves
like (\ref{label2.14}).

The particles with mass $M_{\rm s} = M_{\bar{s}} =
8\sqrt{\alpha_0}/\beta^2$ are single solitons and antisolitons, which
masses do not contain quantum corrections \cite{RD75}. The single
solitons and antisolitons are described by (\ref{label1.6}). The
soliton--soliton and the soliton--antisoliton states read
\cite{SK62}--\cite{RJ75}
\begin{eqnarray}\label{label5.2}
\vartheta_{\rm ss}(x^0,x^1) &=&
\frac{4}{\beta}\,\tan^{-1}\Bigg(u\,\frac{\sinh(\sqrt{\alpha_0}\gamma
x^1)}{\cosh(\sqrt{\alpha_0}\gamma u x^0)}\Bigg),\nonumber\\
\vartheta_{\rm s{\bar{\rm s}}}(x^0,x^1) &=&
\frac{4}{\beta}\,\tan^{-1}\Bigg(\frac{1}{u}
\frac{\sinh(\sqrt{\alpha_0}\gamma u x^0)}{\cosh(\sqrt{\alpha_0}\gamma
x^1)}\Bigg),
\end{eqnarray}
where $\gamma = 1/\sqrt{1 - u^2}$ is the Lorentz factor.

The total energies of these soliton--soliton and soliton-antisoliton
states are equal to $E = 2M_{\rm s}\gamma$ \cite{SK62,RJ75}.

The particles with mass $M^{(n)}_{\rm br} = 2M_{\rm s}\sin \nu_n$ are
the {\it breather} solutions.  Breathers describe soliton--antisoliton
bound states \cite{RD75}. In the rest frame the classical solution
corresponding to the $n$th quantum state reads \cite{AS73,RD75}
\begin{eqnarray}\label{label5.3}
\vartheta^{(n)}_{\rm br}(x^0,x^1) =
\frac{4}{\beta}\,\tan^{-1}\Bigg(\tan\nu_n
\frac{\sin(\sqrt{\alpha_0}x^0\cos \nu_n)}{\cosh(\sqrt{\alpha_0}
x^1\sin\nu_n)}\Bigg).
\end{eqnarray}
As has been shown by Dashen, Hasslacher and Neveu \cite{RD75} small
quantum fluctuations around a one--soliton solution lead to a change
of the soliton (antisoliton) mass as follows
\begin{eqnarray}\label{label5.4}
M_{\rm s} = M_{\bar{\rm s}} = \frac{8\sqrt{\alpha_0}}{\beta^2} -
\frac{\sqrt{\alpha_0}}{\pi} = \frac{8\sqrt{\alpha_0}}{\tilde{\beta}^2},
\end{eqnarray}
where we have denoted
\begin{eqnarray}\label{label5.5}
\tilde{\beta}^2 = \frac{\beta^2}{\displaystyle 1 - \beta^2/8\pi}.
\end{eqnarray}
The masses of breathers are then changed as $M^{(n)}_{\rm br} =
2M_{\rm s}\,\sin \tilde{\nu}_n$, where $\tilde{\nu}_n =
n\tilde{\beta}^2/16$ with $n = 1, 2,\ldots, 8\pi/\tilde{\beta}^2$
\cite{RD75} and $M_{\rm s}$ given by (\ref{label5.4}).

This contribution of quantum fluctuations to the soliton (antisoliton)
mass has been obtained in \cite{RD75} for $\beta^2 < 8\pi$. At
$\beta^2 = 8\pi$ formula (\ref{label5.5}) predicts a singularity.

However, according to Zamolodchikov and Zamolodchikov
\cite{AZ79}:''{\it The singularity of the sine--Gordon theory at
$\beta^2 = 8\pi$ $\ldots$ scarcely means the failure of the theory
with $\beta^2 \ge 8\pi$, but rather indicates a lack of
superrenormalizability property and suggests that another
renormalization prescription is necessary at $\beta^2 \ge 8\pi$}.''

\section{Quantum fluctuations around classical solutions, 
renormalization and soliton--soliton scattering} 
\setcounter{equation}{0}

\hspace{0.2in} The non--perturbative renormalization of the
sine--Gordon model has been carried out in \cite{FI3} (see also
Appendix to this paper). We apply this renormalization procedure to
the calculation of the contribution of quantum fluctuations around a
soliton (antisoliton) solution. The result can be treated as a
continuation of the theory to the region of coupling constants with
$\beta^2 > 8\pi$. We start with the partition function
\begin{eqnarray}\label{label6.1}
Z_{\rm SG} &=& \int {\cal D}\vartheta\,\exp\Big\{i\int
d^2x\,\Big[\frac{1}{2}\,\partial_{\mu}\vartheta(x)
\partial^{\mu}\vartheta(x) +
\frac{\alpha_0}{\beta^2}\,(\cos\beta\vartheta(x) -
1)\Big]\Big\}=\nonumber\\ &=&\int {\cal D}\vartheta\,\exp\Big\{i\int
d^2x\,{\cal L}[\vartheta(x)]\Big\}.
\end{eqnarray}
Following Dashen, Hasslacher and Neveu \cite{RD75} we treat the
fluctuations of the sine--Gordon field $\vartheta(x)$ around the
classical solution $\vartheta(x) = \vartheta_{\rm c\ell}(x) +
\varphi(x)$, where $\vartheta_{\rm c\ell}(x)$ is any classical
solution, satisfying the equations of motion (\ref{label1.5}), and
$\varphi(x)$ is the fluctuating field.

Substituting $\vartheta(x) = \vartheta_{\rm c\ell}(x) + \varphi(x)$
into the exponent of the integrand of (\ref{label6.1}) and using the
equations of motion (\ref{label1.5}) for the classical solution
$\vartheta_{\rm c\ell}(x)$ we get
\begin{eqnarray}\label{label6.2}
Z_{\rm SG} &=& \exp\Big\{i\int d^2x\,{\cal L}[\vartheta_{\rm
c\ell}(x)]\Big\}\nonumber\\ &&\times\,\int {\cal
D}\varphi\,\exp\Big\{i\int
d^2x\,\Big[\frac{1}{2}\,\partial_{\mu}\varphi(x)
\partial^{\mu}\varphi(x) +
\frac{\alpha_0}{\beta}\,\sin\beta\vartheta_{\rm
c\ell}(x)\,\varphi(x)\nonumber\\ && +
\frac{\alpha_0}{\beta^2}\,(\cos(\beta\vartheta_{\rm c\ell}(x) +
\beta\varphi(x)) - \cos\beta\vartheta_{\rm c\ell}(x))\Big]\Big\}.
\end{eqnarray}
In the Gaussian approximation \cite{RD75} the integrand reads
\begin{eqnarray}\label{label6.3}
Z_{\rm SG} &=& \exp\Big\{i\int d^2x\,{\cal L}[\vartheta_{\rm
c\ell}(x)]\Big\}\nonumber\\ &&\times\,\int {\cal
D}\varphi\,\exp\Big\{i\int
d^2x\,\Big[\frac{1}{2}\,\partial_{\mu}\varphi(x)
\partial^{\mu}\varphi(x) -
\frac{1}{2}\,\alpha_0\,\cos\beta\vartheta_{\rm
c\ell}(x)\,\varphi^2(x)\Big]\Big\}.
\end{eqnarray}
The exponent of the integral over $\varphi(x)$ coincides with that in
Eq.(3.4) of Ref.\cite{RD75}.  Integrating over $\varphi(x)$ we obtain
\begin{eqnarray}\label{label6.4}
Z_{\rm SG} &=& \frac{1}{\sqrt{{\rm Det}( \Box +
\alpha_0\,\cos\beta\vartheta_{\rm c\ell})}}\,\exp\Big\{i\int
d^2x\,{\cal L}[\vartheta_{\rm c\ell}(x)]\Big\} = \nonumber\\
&=&\exp\Big\{i\int d^2x\,{\cal L}_{\rm eff}[\vartheta_{\rm
c\ell}(x)]\Big\},
\end{eqnarray}
where the effective Lagrangian ${\cal L}_{\rm eff}[\vartheta_{\rm
c\ell}(x)]$ is defined by
\begin{eqnarray}\label{label6.5}
{\cal L}_{\rm eff}[\vartheta_{\rm c\ell}(x)] = {\cal L}[\vartheta_{\rm
c\ell}(x)] + i\,\frac{1}{2}\,\Big\langle x\Big|{\ell n}\Big(1 +
\frac{\alpha_0}{ \Box + \alpha_0}(\cos\beta\vartheta_{\rm c\ell}(x) -
1)\Big)\Big|x\Big\rangle.
\end{eqnarray}
The wave functions $|x\rangle$ are normalized by $\langle x|y\rangle =
\delta^{(2)}(x - y)$ \cite{FI1,RB96}.

The first order correction to ${\cal L}[\vartheta_{\rm c\ell}(x)]$ is
equal to
\begin{eqnarray}\label{label6.6}
\hspace{-0.3in}&&{\cal L}^{(1)}[\vartheta_{\rm c\ell}(x)] =
-\,\frac{1}{2}\int \frac{d^2k}{(2\pi)^2i}\,\frac{\alpha_0}{\alpha_0 -
k^2 - i\,0}\,(\cos\beta\vartheta_{\rm c\ell}(x) - 1) =\nonumber\\
\hspace{-0.3in}&&= -
\frac{\alpha_0}{8\pi}\int^{\Lambda^2}_0\frac{dk^2_E}{\alpha_0 +
k^2_E}\,(\cos\beta\vartheta_{\rm c\ell}(x) - 1) = -
\frac{\alpha_0}{8\pi}\,{\ell n}\Big(\frac{\Lambda^2}{\alpha_0}\Big)\,
(\cos\beta\vartheta_{\rm c\ell}(x) - 1),
\end{eqnarray}
where $\Lambda$ is the ultra--violet cut--off. We carried out the Wick
rotation to the Euclidean momentum space $d^2k = id^2k_E$ and $k^2 =
-k^2_E$.

For the effective Lagrangian ${\cal L}_{\rm eff}[\vartheta_{\rm
c\ell}(x)]$ we obtain
\begin{eqnarray}\label{label6.7}
\hspace{-0.3in}{\cal L}_{\rm eff}[\vartheta_{\rm c\ell}(x)] &=&
\frac{1}{2}\,\partial_{\mu}\vartheta_{\rm
c\ell}(x)\partial^{\mu}\vartheta_{\rm c\ell}(x) +
\frac{\alpha_0}{\beta^2}\,\Big[1 - \frac{\beta^2}{8\pi}\,{\ell
n}\Big(\frac{\Lambda^2}{\alpha_0}\Big)\Big]\, (\cos\beta\vartheta_{\rm
c\ell}(x) - 1).
\end{eqnarray}
Now we have to renormalize the coupling constant $\alpha_0$ in oder to
remove the ultra--violet cut--off $\Lambda$. The coupling constant
$\alpha(M)$ renormalized at the normalization scale $M$ is defined by
$\alpha(M) = Z^{-1}_1(\beta, M, \alpha_0; \Lambda) \,\alpha_0$ (see
Appendix). The renormalization constant $Z_1(\beta, M, \alpha_0;
\Lambda)$ is equal to ({\rm A}.13)
\begin{eqnarray}\label{label6.8}
Z_1(\beta, M, \alpha_0; \Lambda) =
\Big(\frac{\Lambda^2}{M^2}\Big)^{\textstyle \beta^2/8\pi} = 1 +
\frac{\beta^2}{8\pi}\,{\ell n}\Big(\frac{\Lambda^2}{M^2}\Big) +
\ldots\,.
\end{eqnarray}
The renormalization of the effective potential in the effective
Lagrangian (\ref{label6.7}) runs as follows. Treating only the
constant factor in front of $(\cos\beta\vartheta_{\rm c\ell}(x) - 1)$
we get
\begin{eqnarray}\label{label6.9}
&&\frac{\alpha_0}{\beta^2}\,\Big[1 - \frac{\beta^2}{8\pi}\,{\ell
n}\Big(\frac{\Lambda^2}{\alpha_0}\Big)\Big] = \frac{\alpha
Z_1}{\beta^2}\,\Big[1 - \frac{\beta^2}{8\pi}\,{\ell
n}\Big(\frac{\Lambda^2}{\alpha Z_1}\Big)\Big] = \nonumber\\ &&=
\frac{\alpha }{\beta^2}\,\Big[1 - \frac{\beta^2}{8\pi}\,{\ell
n}\Big(\frac{\Lambda^2}{\alpha}\Big) + (Z_1 - 1)\Big] =
\frac{\alpha }{\beta^2}\,\Big[1 - \frac{\beta^2}{8\pi}\,{\ell
n}\Big(\frac{\Lambda^2}{\alpha}\Big) + \frac{\beta^2}{8\pi}\,{\ell
n}\Big(\frac{\Lambda^2}{M^2}\Big)\Big] = \nonumber\\ &&=
\frac{\alpha}{\beta^2}\, \Big[1 - \frac{\beta^2}{8\pi}\,{\ell
n}\Big(\frac{M^2}{\alpha}\Big)\Big],
\end{eqnarray}
where we have dropped the terms of order of $O(\beta^4)$. 

Thus, the renormalized effective Lagrangian of the sine--Gordon model
reads
\begin{eqnarray}\label{label6.10}
\hspace{-0.3in}{\cal L}^{(r)}_{\rm eff}[\vartheta_{\rm c\ell}(x)] &=&
\frac{1}{2}\,\partial_{\mu}\vartheta_{\rm
c\ell}(x)\partial^{\mu}\vartheta_{\rm c\ell}(x) +
\frac{\alpha}{\beta^2}\,\Big[1 - \frac{\beta^2}{8\pi}\,{\ell
n}\Big(\frac{M^2}{\alpha}\Big)\Big]\, (\cos\beta\vartheta_{\rm
c\ell}(x) - 1),
\end{eqnarray}
where we have denoted $\alpha = \alpha(M)$.

The non--perturbative correction, caused by quantum fluctuations
around a classical solution, to the effective potential of the
sine--Gordon model can be written as
\begin{eqnarray}\label{label6.11}
\hspace{-0.3in}{\cal L}^{(r)}_{\rm eff}[\vartheta_{\rm c\ell}(x)] &=&
\frac{1}{2}\,\partial_{\mu}\vartheta_{\rm
c\ell}(x)\partial^{\mu}\vartheta_{\rm c\ell}(x) +
\frac{\alpha}{\beta^2}\,\Big(\frac{\alpha}{M^2}\Big)^{\textstyle
\beta^2/8\pi}\, (\cos\beta\vartheta_{\rm c\ell}(x) - 1).
\end{eqnarray}
This agrees with our expression for the energy density of the ground
state of the sine--Gordon model (\ref{label3.10}), where the
ultra--violet cut--off is equal to the renormalized mass of the
sine--Gordon quanta, $\Lambda = \sqrt{\alpha}$.

The most convenient choice of the renormalization point is $M =
\alpha(M)$. This yields
\begin{eqnarray}\label{label6.12}
{\cal L}^{(r)}[\vartheta_{\rm c\ell}(x)] =
\frac{1}{2}\,\partial_{\mu}\vartheta_{\rm
c\ell}(x)\partial^{\mu}\vartheta_{\rm s}(x) +
\frac{\alpha}{\beta^2}\,(\cos\beta\vartheta_{\rm c\ell}(x) - 1).
\end{eqnarray}
Since we have not specified the classical solution, our result is
valid for quantum corrections around an arbitrary classical solution
of the sine--Gordon model.  Our result of the calculation of the
quantum fluctuations agrees with that carried out by Korepin, Kulish
and Faddeev \cite{VK75}.

According to the renormalized Lagrangian (\ref{label6.10}) the soliton
(antisoliton) mass is equal to $M_{\rm s} = M_{\bar{\rm s}} =
8\sqrt{\alpha}/\beta^2$. The masses of breathers would be changed as
follows $M^{(n)}_{\rm br} = (16\sqrt{\alpha}/\beta^2)\,\sin\nu_n$ with
$\nu_n = n\beta^2/16$ and $n = 1,2,\ldots, 8\pi/\beta^2$.

Hence, quantum fluctuations, calculated with the renormalization
prescription expounded above, do not lead to the appearance of a
singular point in the sine--Gordon model and allow the continuation of
the theory to the region $\beta^2 \ge 8\pi$ as has been suspected by
Zamolodchikov and Zamolodchikov \cite{AZ79}.

As has been pointed out in \cite{FI1} for $\beta^2 > 8\pi$ the
1+1--dimensional world is populated mainly by solitons and
antisolitons. Breather states are prohibited for $\beta^2 >
8\pi$. This agrees with the assertion by Zamolodchikov and
Zamolodchikov \cite{AZ79}, which reads in our notation: ``{\it At
$\beta^2 > 8\pi$ all bound states including the ``elementary''
particle of the sine--Gordon Lagrangian (\ref{label1.2}) become
unbound. Thus, at $\beta^2 \ge 8\pi$ the spectrum contains solitons
and antisolitons only.}''

The phase shift for soliton--soliton scattering has been calculated by
Weisz in dependence on the rapidity difference $\theta$ and the
sine--Gordon coupling constant $\lambda > 1$ \cite{PW77}. For $-\infty
< {\cal R}e \theta < + \infty$ and $|{\cal I}m \theta| < {\rm
min}[\pi, \lambda \pi ]$ the integral representation for the phase
shift reads
\begin{eqnarray}\label{label6.13}
\delta_{\rm ss}(\theta) =
\frac{1}{2}\int^{\infty}_0\frac{dt}{t}\,\frac{\displaystyle
\sin\Big(\frac{\theta t}{\pi}\Big)\sinh\Big(\frac{1}{2} (\lambda -
1)\,t\Big)}{\displaystyle \sin\Big(\frac{1}{2}\,\lambda\,t\Big)
\cosh\Big(\frac{1}{2}\,t\Big)}
\end{eqnarray}
and ``{\it exhibits the absence of physical bound states for $\lambda
> 1$.}'' \cite{PW77}. In our renormalization procedure expounded above
$\lambda = \beta^2/8\pi > 1$.

Thus, the phase shift $\delta_{\rm ss}(\theta)$, defined by
(\ref{label6.13}), should describe soliton--soliton scattering for the
sine--Gordon coupling constants obeying the constraint $\beta^2 >
8\pi$.

The absence of contributions from soliton--antisoliton bound states to
the phase shift $\delta_{\rm ss}(\theta)$ of soliton--soliton
scattering for $\beta^2 > 8\pi$ agrees with conclusions by
Zamolodchikov and Zamolodchikov \cite{AZ79} and ours, concerning the
population of the 1+1--dimensional world by only solitons and
antisolitons for $\beta^2 > 8\pi$.

\section{Conclusion}
\setcounter{equation}{0}

\hspace{0.2in} We have shown that the vacuum energy density of the
ground state of the sine--Gordon model is bounded from below even for
$\beta^2 > 8\pi$. We have found some unconvincing assumptions of
Coleman's proof. These are (i) the parameter $\gamma_0$, normalizing to
zero the classical potential energy of the sine--Gordon model, has
been assumed additively renormalizable and set finite after
renormalization, (ii) the renormalized Hamiltonian has been found
depending on the infrared cut--off $\mu$ with divergent contributions
in the limit $\mu \to 0$ and (iii) the vacuum energy density
(\ref{label2.21}) calculated by Coleman is equal to ${\cal E}_{\rm
vac}(M) = - \infty$ in the infrared limit $\mu \to 0$ for any finite
scale $M$ and coupling constant $\beta$, whereas the sine--Gordon
model is well--defined in the infrared limit $\mu \to 0$, see the
Appendix.

Our direct calculation of the vacuum energy density is
non--perturbative and exact. We have shown explicitly that the vacuum
energy density of the sine--Gordon model can never be a negative
quantity if the potential energy is normalized to zero as it is done
at the classical level.

Summarizing the obtained results we can conclude that in the region of
coupling constants, obeying the constraint $\beta^2 > 8\pi$, the
sine--Gordon model can be treated well. For the coupling constants
$\beta^2 > 8\pi$ the sine--Gordon model describes only solitons and
antisolitons without breathers. The amplitudes of scattering of
soliton by soliton and soliton by antisoliton are well--defined for
$\beta^2 > 8\pi$ without soliton--antisoliton bound state
contributions to the intermediate states. In our approach the
sine--Gordon model for coupling constants $\beta^2 > 8\pi$ is a
quantum field theory with Wightman's observables defined on the test
functions from ${\cal S}_0(\mathbb{R}^2)$
\cite{FI2,FI8,FI7}. Therefore, it does not contradict Coleman's
theorem, asserting the absence of spontaneously broken continuous
symmetry in quantum field theories with Wightman's observables defined
on the test functions from ${\cal S}(\mathbb{R}^2)$.

\section*{Acknowledgement}

\hspace{0.2in} This work was supported in part by Fonds zur
F\"orderung der Wissenschaftlichen Forschung P11387--PHY.

\newpage

\section*{Appendix.  Non--perturbative 
renormalizability of the sine--Gordon model}

\hspace{0.2in} As has been shown in \cite{FI3} the massless Thirring
model with non--vanishing external sources is equivalent to the
sine--Gordon model, where the mass of Thirring fermion fields $m$ is
considered as an external source $\sigma(x) = - m$ for the scalar
fermion density $\bar{\psi}(x)\psi(x)$. Therefore, the properties of
non--perturbative renormalizability of the massless Thirring model
investigated in \cite{FI3} should be fully extended to the
sine--Gordon (SG) model.

The generating functional of Green functions in the SG  model we define
as
$$
Z_{\rm SG}[J] = \int {\cal D}\vartheta\,\exp\,i\int
d^2x\,\Big\{\frac{1}{2}\,\partial_{\mu}\vartheta(x)\partial^{\mu}
\vartheta(x) + \frac{\alpha_0}{\beta^2}\,(\cos\beta\vartheta(x) - 1) +
\vartheta(x)J(x)\Big\},\eqno({\rm A}.1)
$$
where $J(x)$ is an external source of the $\vartheta(x)$--field.

The Lagrangian of the SG model is invariant under the transformations
$$
\vartheta(x) \to \vartheta\,'(x) = \vartheta(x) + \frac{2\pi
n}{\beta},\eqno({\rm A}.2)
$$
where $n$ is an integer number running over $n = 0,\pm 1,\pm
2,\ldots$.  In order to get the generating functional $Z_{\rm SG}[J]$
invariant under the transformations ({\rm A}.2) it is sufficient to
restrict the class of functions describing the external source of the
$\vartheta$--field and impose the constraint \cite{FI2}
$$
\int d^2x\,J(x) = 0.\eqno({\rm A}.3)
$$
Non--perturbative renormalizability of the SG model we understand as a
possibility to remove all divergences by renormalizing the coupling
constant $\alpha_0$. Indeed, since the coupling constant $\beta$ is
related to the coupling constant of the Thirring model $g$
\cite{FI1,FI3} which is unrenormalized $g_0 = g$, so the coupling
constant $\beta$ should possess the same property, i.e. $\beta_0 =
\beta$. Hence, only the coupling constant $\alpha_0$ should undergo
renormalization.

The Lagrangian of the SG model written in terms of {\it bare}
quantities reads
$$
{\cal L}_{\rm SG}(x) =
\frac{1}{2}\partial_{\mu}\vartheta_0(x)\partial^{\mu}\vartheta_0(x) +
\frac{\alpha_0}{\beta^2}\,(\cos\beta\vartheta_0(x) - 1).\eqno({\rm
A}.4)
$$
Since $\beta$ is the unrenormalized coupling constant, the field
$\vartheta_0(x)$ should be also unrenormalized, $\vartheta_0(x) =
\vartheta(x)$. This means that there is no renormalization of the wave
function of the $\vartheta$--field.  As a result the Lagrangian ${\cal
L}_{\rm SG}(x)$ of the SG model in terms of renormalized quantities
can be written by
$$
{\cal L}_{\rm SG}(x) =
\frac{1}{2}\partial_{\mu}\vartheta(x)\partial^{\mu}\vartheta(x) +
\frac{\alpha}{\beta^2}\,(\cos\beta\vartheta(x) - 1)
+ (Z_1 -
1)\,\frac{\alpha}{\beta^2}\,(\cos\beta\vartheta(x) -
1) =
$$
$$
= \frac{1}{2}\partial_{\mu}\vartheta(x)\partial^{\mu}\vartheta(x) +
Z_1\,\frac{\alpha}{\beta^2}\,(\cos\beta\vartheta(x) - 1),\eqno({\rm
A}.5)
$$
where $Z_1$ is the renormalization constant of the coupling constant
$\alpha$. The renormalized coupling constant $\alpha$ is related to
the {\it bare} one by the relation 
$$
\alpha = Z^{-1}_1\,\alpha_0.\eqno({\rm A}.6)
$$
Renormalizability of the SG model as well as the Thirring model we
understand as the possibility to replace the ultra--violet cut--off
$\Lambda$ by another finite scale $M$ by means of the renormalization
constant $Z_1$ in the limit $\mu \to 0$. According to the general
theory of renormalizations \cite{JC84} $Z_1$ should be a function of
the coupling constants $\beta$, $\alpha$, the infrared cut--off $\mu$,
the ultra--violet cut--off $\Lambda$ and a finite scale $M$:
$$
Z_1 = Z_1(\beta,\alpha, M; \mu, \Lambda).\eqno({\rm
A}.7)
$$
Now let us proceed to the evaluation of the generating functional
({\rm A}.1). For this aim expand the integrand of the generating
functional $Z_{\rm SG}[J]$ in powers of
$\alpha_0\cos\beta\vartheta(x)$. This gives
$$
Z_{\rm SG}[J] = \lim_{\textstyle \mu \to 0}e^{\textstyle -i\int
d^2x\,{\displaystyle
\frac{\alpha_0}{\beta^2}}}\sum^{\infty}_{n=0}
\frac{i^n}{n!}\,\Bigg(\frac{\alpha_0}{\beta^2}\Bigg)^n
\prod^n_{i=1}\int d^2x_i
$$
$$
\times \int {\cal
D}\vartheta\prod^n_{i=1}\cos\beta\vartheta(x_i)\,\exp\, i\int
d^2x\,\Big\{\frac{1}{2}\,\partial_{\mu}\vartheta(x)
\partial^{\mu}\vartheta(x) - \frac{1}{2}\,\mu^2\vartheta^2(x) +
\vartheta(x)J(x)\Big\}.\eqno({\rm A}.8)
$$
The integration over the $\vartheta$--field can be carried out
explicitly and we get 
$$
Z_{\rm SG}[J] = \lim_{\textstyle \mu \to 0}e^{\textstyle -i\int
d^2x\,{\displaystyle \frac{\alpha_0}{\beta^2}}}\sum^{\infty}_{n=0}
\frac{i^n}{n!}\,\Bigg(\frac{\alpha_0}{2\beta^2}\Bigg)^n
\sum^n_{p=0}\frac{n!}{(n-p)!\,p!}\prod^{n-p}_{j=1}\prod^p_{k=1}\int
d^2x_j d^2y_k
$$
$$
\times\,\exp\,\Big\{\frac{1}{2}\,n\,\beta^2 i \Delta(0;\mu) +
\beta^2\sum^{n-p}_{j < k}i\Delta(x_j - x_k;\mu) + \beta^2\sum^{p}_{j <
k}i \Delta(y_j - y_k;\mu)
$$
$$
 - \beta^2\sum^{n-p}_{j = 1}\sum^{p}_{k = 1}i\Delta(x_j -
y_k;\mu)\Big\}\,\exp\,\Big\{\int d^2x\,\beta\,[\sum^{n-p}_{j =
1}i\Delta(x_j - x;\mu) - \sum^{p}_{k = 1}i\Delta(y_j - x;\mu)]\,J(x)
$$
$$
 + \int\!\!\!\int d^2x\,d^2y\,\frac{1}{2}\,J(x)\,i\Delta(x -
y;\mu)\,J(y)\Big\},\eqno({\rm A}.9)
$$
where the causal Green functions $\Delta(x-y;\mu)$ and $\Delta(0;\mu)$
are defined by \cite{FI3,FI2}
$$
\Delta(x-y;\mu) = i\theta(x^0 - y^0)D^{(+)}(x-y;\mu) + i\theta(y^0 -
x^0)D^{(-)}(y-x;\mu) = - \frac{i}{4\pi}{\ell n}[-\mu^2(x-y)^2 + i0],
$$
$$
\Delta(0;\mu) = \frac{i}{4\pi}\,{\ell
n}\Big(\frac{\Lambda^2}{\mu^2}\Big).
$$
Taking the limit $\mu \to 0$ we reduce the r.h.s. of ({\rm A}.9) to
the form
$$
Z_{\rm SG}[J] = e^{\textstyle -i\int d^2x\,{\displaystyle
\frac{\alpha_0}{\beta^2}}}\sum^{\infty}_{p=0}
\frac{(-1)^p}{(p!)^2}\,
\Bigg(\frac{\alpha_0}{2\beta^2}\Bigg)^{2p}
\prod^{p}_{j=1}\int\!\!\!\int d^2x_j d^2y_j\,
\Bigg[\Bigg(\frac{M^2}{\Lambda^2}\Bigg)^{\textstyle
\beta^2/8\pi}\Bigg]^{\!2p}
$$
$$
\times\exp\Big\{\frac{\beta^2}{4\pi}\sum^p_{j < k} \Big({\ell n}[ -
M^2(x_j - x_k)^2 + i0] + {\ell n}[ - M^2(y_j - y_k)^2 + i0]\Big)
$$
$$
 - \frac{\beta^2}{4\pi}\sum^p_{j=1}\sum^p_{k=1}{\ell n}[ - M^2(x_j -
y_k)^2 + i0]\Big\}\,\exp\Big\{\frac{\beta}{4\pi}\int d^2x\,\sum^p_{j =
1} {\ell n}\Bigg[\frac{(x_j - x)^2 + i0 }{(y_j - x)^2 +
i0}\Bigg]\,J(x)
$$
$$
 + \frac{1}{8\pi}\int\!\!\!\int d^2x\,d^2y\,J(x)\,{\ell
n}[-M^2(x-y)^2+ i0]\,J(y)\Big\}
$$
$$
\times\lim_{\textstyle \mu \to 0} \exp\Big\{ -\frac{1}{4\pi}\,{\ell
n}\Big(\frac{M}{\mu}\Big) \Big(\int
d^2x\,J(x)\Big)^{\!\!2}\Big\}.\eqno({\rm A}.10)
$$
Due to the constraint ({\rm A}.3) the generating functional $Z_{\rm
SG}[J]$ does not depend on the infrared cut--off $\mu$.  Using ({\rm
A}.3) we get
$$
Z_{\rm SG}[J] = e^{\textstyle -i\int
d^2x\,{\displaystyle
\frac{\alpha_0}{\beta^2}}}\sum^{\infty}_{p=0} \frac{(-1)^p}{(p!)^2}\,
\Bigg(\frac{\alpha_0}{2\beta^2}\Bigg)^{2p}
\prod^{p}_{j=1}\int\!\!\!\int d^2x_j d^2y_j\,
\Bigg[\Bigg(\frac{M^2}{\Lambda^2}\Bigg)^{\textstyle
\beta^2/8\pi}\Bigg]^{\!2p}
$$
$$
\times\exp\Big\{\frac{\beta^2}{4\pi}\sum^p_{j < k} \Big({\ell n}[ -
M^2(x_j - x_k)^2 + i0] + {\ell n}[ - M^2(y_j - y_k)^2 + i0]\Big)
$$
$$
 -
\frac{\beta^2}{4\pi}\sum^p_{j=1}\sum^p_{k=1}{\ell n}[ - M^2(x_j -
y_k)^2 + i0]\Big\}\,\exp\Big\{\frac{\beta}{4\pi}\int d^2x\,\sum^p_{j = 1} {\ell
n}\Bigg[\frac{(x_j - x)^2 + i0}{(y_j - x)^2 + i0}\Bigg]\,J(x)
$$
$$
 + \frac{1}{8\pi}\int\!\!\!\int d^2x\,d^2y\,J(x)\,{\ell n}[-M^2(x-y)^2
+ i0]\,J(y)\Big\}.\eqno({\rm A}.11)
$$
Passing to a renormalized constant $\alpha$, $\alpha_0 = Z_1\alpha$,
we recast the r.h.s. of ({\rm A}.11) into the form
$$
Z_{\rm SG}[J] = e^{\textstyle -i\int d^2x\,{\displaystyle
Z_1\,\frac{\alpha}{\beta^2}}}\sum^{\infty}_{p=0} \Bigg[Z_1
\Bigg(\frac{M^2}{\Lambda^2}\Bigg)^{ \!\textstyle
\beta^2/8\pi}\Bigg]^{\!2p} \frac{(-1)^p}{(p!)^2}\,
\Bigg(\frac{\alpha}{2\beta^2}\Bigg)^{2p}
\prod^{p}_{j=1}\int\!\!\!\int d^2x_j d^2y_j\,
$$
$$
\times\exp\Big\{\frac{\beta^2}{4\pi}\sum^p_{j < k} \Big({\ell
n}[ - M^2(x_j - x_k)^2 + i0] + {\ell n}[ - M^2(y_j - y_k)^2 + i0]\Big)
$$
$$
 - \frac{\beta^2}{4\pi}\sum^p_{j=1}\sum^p_{k=1}{\ell n}[ - M^2(x_j -
y_k)^2 + i0]\Big\}\,\exp\Big\{\frac{\beta}{4\pi}\int d^2x\,\sum^p_{j =
1} {\ell n}\Bigg[\frac{(x_j - x)^2 + i0}{(y_j - x)^2 + i0}\Bigg]\,J(x)
$$
$$
+ \frac{1}{8\pi}\int\!\!\!\int d^2x\,d^2y\,J(x)\,{\ell n}[-M^2(x-y)^2
+ i0]\,J(y)\Big\}.\eqno({\rm A}.12)
$$
Setting 
$$
Z_1 = \Bigg(\frac{\Lambda^2}{M^2}\Bigg)^{ \!\textstyle
\beta^2/8\pi}\eqno({\rm A}.13)
$$
we are left with the dependence of the generating functional $Z_{\rm
SG}[J]$ on the ultra--violet cut--off $\Lambda$ only in the
insignificant constant factor
$$
Z_{\rm SG}[J] = e^{\textstyle -i\int d^2x\,{\displaystyle
\frac{\alpha}{\beta^2}}\,\Bigg(\frac{\textstyle \Lambda^2}{\textstyle
M^2}\Bigg)^{ \!\textstyle
\beta^2/8\pi}}\sum^{\infty}_{p=0}\frac{(-1)^p}{(p!)^2}\,
\Bigg(\frac{\alpha}{2\beta^2}\Bigg)^{2p} \prod^{p}_{j=1}\int\!\!\!\int
d^2x_j d^2y_j\,
$$
$$
\times\exp\Big\{\frac{\beta^2}{4\pi}\sum^p_{j < k} \Big({\ell n}[ -
M^2(x_j - x_k)^2 + i0] + {\ell n}[ - M^2(y_j - y_k)^2 + i0]\Big)
$$
$$
 - \frac{\beta^2}{4\pi}\sum^p_{j=1}\sum^p_{k=1}{\ell n}[ - M^2(x_j -
y_k)^2 + i0]\Big\}\,\exp\Big\{\frac{\beta}{4\pi}\int d^2x\, \sum^p_{j
= 1} {\ell n}\Bigg[\frac{(x_j - x)^2 + i0}{(y_j - x)^2 +
i0}\Bigg]\,J(x)
$$
$$
 + \frac{1}{8\pi}\int\!\!\!\int d^2x\,d^2y\,J(x)\,{\ell n}[-M^2(x-y)^2
+ i0]\,J(y)\Big\}.\eqno({\rm A}.14)
$$
The generating functional ({\rm A}.14) is expressed in terms of the
renormalized constant $\alpha$, the constant $\beta$ and the finite
scale $M$. The ultra--violet cut--off $\Lambda$ enters only in the
insignificant constant factor, which does not affect the result of the
evaluation of correlation functions. This factor can be removed by
redefinition of the path--integral measure of the generating
functional $Z_{\rm SG}[J]$.

Thus, the generating functional $Z_{\rm SG}[J]$ ({\rm A}.14) can be
applied to the evaluation of any renormalized correlation function of
the SG model. This testifies the complete non--perturbative
renormalizability of the SG model.

Using ({\rm A}.14) we evaluate the partition function $Z_{\rm
SG}[0]$. It is equal to
$$
Z_{\rm SG}[0] = e^{\textstyle -i\int d^2x\,{\displaystyle
\frac{\alpha}{\beta^2}}\,\Bigg(\frac{\textstyle \Lambda^2}{\textstyle
M^2}\Bigg)^{ \!\textstyle
\beta^2/8\pi}}\sum^{\infty}_{p=0}\frac{(-1)^p}{(p!)^2}\,
\Bigg(\frac{\alpha}{2\beta^2}\Bigg)^{2p} \prod^{p}_{j=1}\int\!\!\!\int
d^2x_j d^2y_j\,
$$
$$
\times\exp\Big\{\frac{\beta^2}{4\pi}\sum^p_{j < k} \Big({\ell n}[ -
M^2(x_j - x_k)^2 + i0] + {\ell n}[ - M^2(y_j - y_k)^2 + i0]\Big)
$$
$$
 - \frac{\beta^2}{4\pi}\sum^p_{j=1}\sum^p_{k=1}{\ell n}[ - M^2(x_j -
y_k)^2 + i0]\Big\}.\eqno({\rm A}.15)
$$
This expression we use for the calculation of the vacuum energy
density of the SG model.

\end{document}